# $^3$He NMR in porous media: Inverse Laplace transformation

R.R. Gazizulin, A.V. Klochkov, V.V. Kuzmin, K.R. Safiullin, M.S. Tagirov, A.N. Yudin

Kazan (Volga region) Federal University, Kremlevskaya, 18, Kazan 420008, Russia

e-mail: g-rasul@yandex.ru, klochkov@gmail.com, murat.tagirov@ksu.ru

For the first time the inverse Laplace transform was applied for analysis of $^3$He relaxation in porous media. It was shown that inverse Laplace transform gives new information about these systems. The uniform-penalty algorithm has been performed to obtain the $^3$He relaxation times distribution in pores of clay sample. It is possible to obtain pores' sizes distribution by using applicable model.

## 1. Introduction

The use of inverse Laplace transform to investigate some relaxation curves is well-known problem in many fields of material research [1-2]. The nuclear magnetic resonance (NMR) in porous media and crystal powder samples is one of the most informative methods for investigating of such systems [3-9]. The measuring longitudinal magnetization relaxation time $T_1$ is important in this investigation. Usually, there is a distribution of the relaxation times $T_1$ in most of the samples. Obtaining the relaxation times distribution from the longitudinal magnetization recovery curve is important for investigating properties of the systems. Various algorithms have been used to analyze experimental relaxation data. Fundamental requirement for such algorithms is the obtained distribution accordance to the properties of the real systems. In our work we have researched the applicability of the uniform-penalty inversion algorithm [10] for investigating properties of the systems by measuring NMR $^3$He relaxation in the clay sample.

The pulse NMR measurements are a very useful tool for understanding various properties of different systems. One of the key parameters that are widely used in pulse NMR is the longitudinal relaxation time $T_1$. This time constant characterizes the process of longitudinal magnetization recovery. This process is commonly described by the following equation:

$$A(t) = A_0\left(1-\exp\left(-\frac{t}{T_1}\right)\right), \quad (1)$$

In most porous systems, for example in clay samples, the diffusion isn't fast enough and we can assume that the molecules are located in the same pore during experiment. In this case the relaxation of whole spin system is the superposition of contributions from the pores with different sizes, which results in $T_1$ distribution. But the experimental data usually can be described by following equation in most cases:

$$A(t) = A_0\left(1-\exp\left(-\left(\frac{t}{T_1}\right)^n\right)\right), \quad (2)$$

Such equation contains all information about $T_1$ distribution. To have a clear understanding of sample properties through NMR measurements it is necessary to obtain reliable inversion of experimental data.

The experimental data represent sums of exponentially decaying components:

$$A(t_i) \approx A_0 + \sum_k f_k(T_1)\exp\left(-\frac{t_i}{T_k}\right), \qquad (3)$$

where $f_k(T_1)$ is the weighting factor.

The uniform-penalty inversion (UPEN) algorithm is one of the most succesfull regularization processes applied to NMR relaxation data. The uniform-penalty algorithm can give sharp lines, not broadened more than conditioned by noise, and in the same distribution it can show wide lines, not broken into several peaks [10]. In general, this algorithm represents a least square minimization routine with adding a penalty factor. A coefficient $C_k$ is multiplied to the penalty factor. The coefficient $C_k$ is iteratively adjusted to ensure a unifromity of the penalty factor. Thus, the function to be minimized is:

$$\sum_i \left(A_0 + \sum_k f_k(T_1)\exp\left(-\frac{t_i}{T_{1k}}\right) - A(t_i)\right)^2 + \sum_k C_k\left(f_{k-1}(T_1) - 2f_k(T_1) + f_{k+1}(T_1)\right)^2, \qquad (4)$$

where the first component is the least minimization routine component and the second component is the penalty factor.

Using this algorithm we have created a program which contains the inverse Laplace transform. For the first time it has been applied to the magnetic relaxation of liquid $^3$He.

## 2. Methods

The simulated data were created for single component $T_1$ process with $T_1$ = 11.9 s. Then real experimental noise pattern was added to simulated mono-exponential curve so that signal-to-noise ratio was the same order of magnitude which we usually obtain in our experiments (Fig.1).

UPEN from simulated mono-exponential recovery data is shown on Fig. 2, where the pick with finite width can be seen. As the noise added, the broadening of the computed peak is observed.

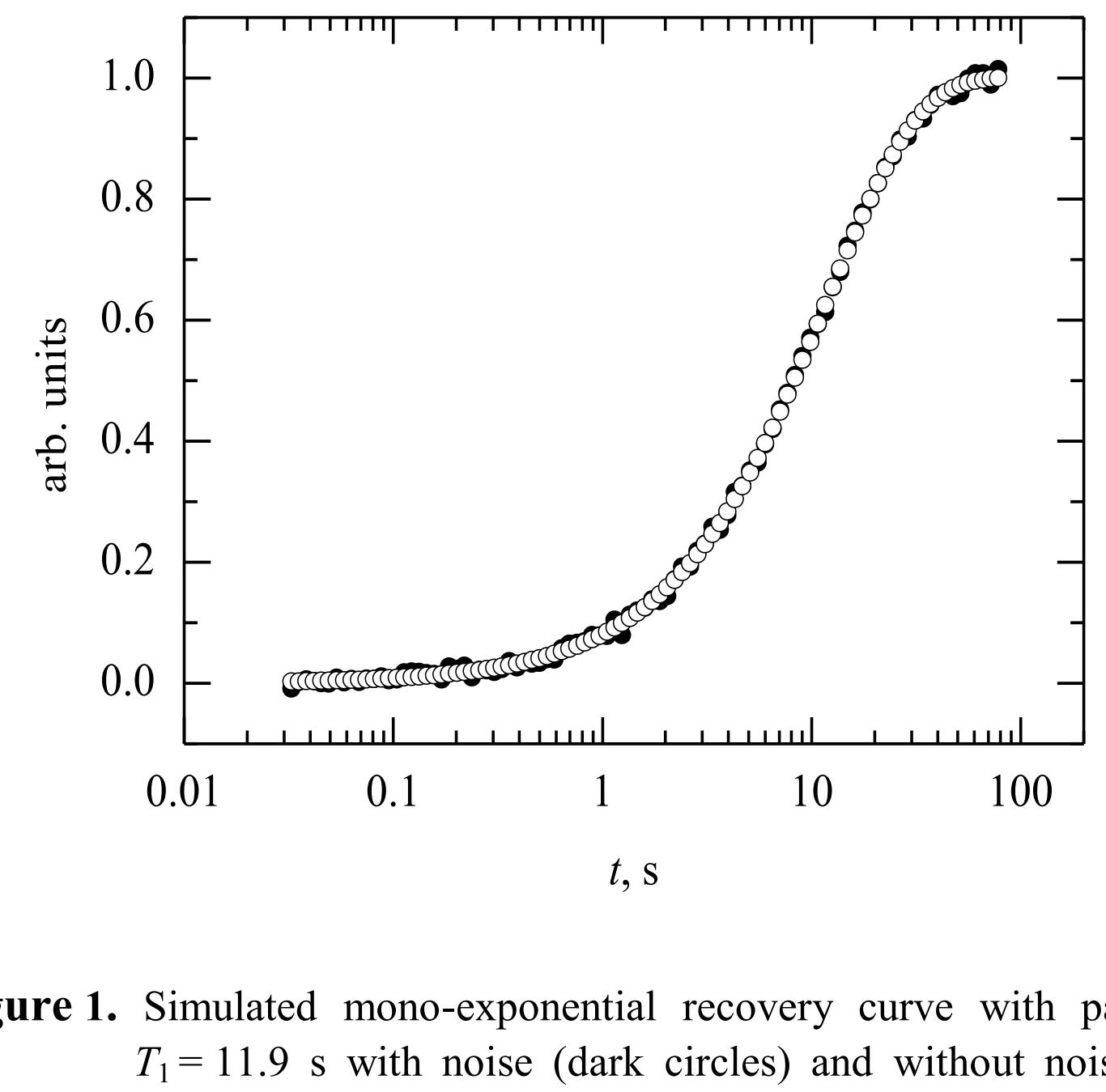

**Figure 1.** Simulated mono-exponential recovery curve with parameter $T_1$ = 11.9 s with noise (dark circles) and without noise (open circles).

To investigate the applicability of the uniform-penalty algorithm for describing properties of real systems by the NMR relaxation measurements of $^3$He, the specially prepared sample of clay filled by $^3$He was used. In this system $T_1$ relaxation data of $^3$He contain information about porosity of the sample and the time distribution probably corresponds to pores' sizes distribution. The clay sample was cut out of clay mineral in the shape of tablet with diameter of 5 mm and height of 3 mm. A hole (diameter of 1.7 mm) has been drilled out in the center, along the tablet axis, forming a calibration cavity (Fig. 3). The relaxation in the calibration cavity has to be mono-exponential and has much bigger contribution to relaxation curve than contribution from $^3$He relaxation in each and any pore. The idea is to obtain the contribution from calibration cavity into the $^3$He relaxation time distribution. By comparison this contribution with simulated mono-exponential distributions obtained by the uniform-penalty algorithm we can make a conclusion about its applicability.

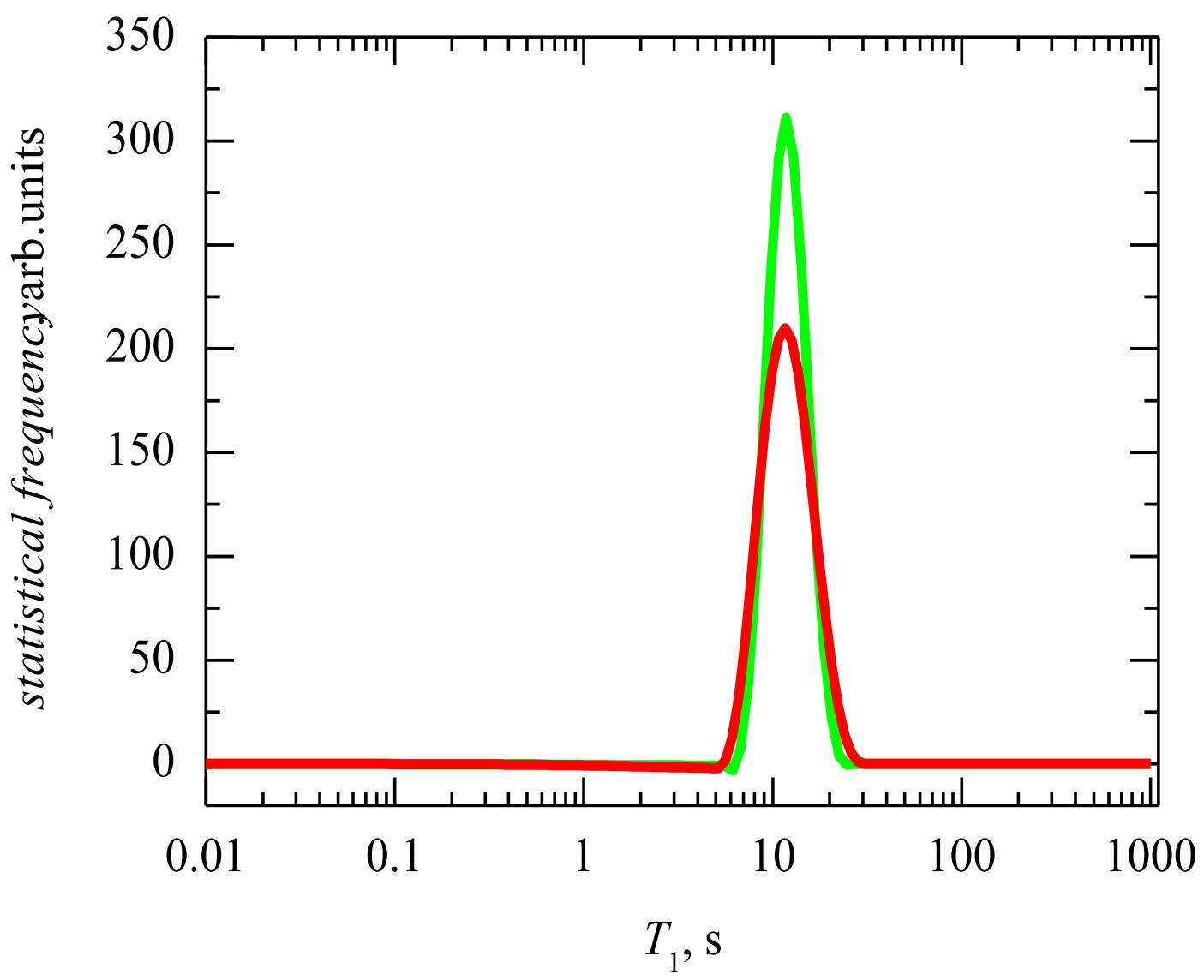

**Figure 2.** UPEN from simulated mono-exponential data without noise (green curve) and UPEN from simulated mono-exponential data with noise (red curve).

The sample was placed in the pyrex tube which was sealed leak tight to the $^3$He gas handling system. On the outer surface of the pyrex tube a NMR coil was mounted. The hand made pulse NMR spectrometer has been used (frequency range 3 – 50 MHz). The pulse NMR spectrometer is equipped by resistive electrical magnet with magnetic field strength up to 10 kOe. The sample has been filled by $^3$He under saturation vapor pressure at the temperature 1.5 K (Fig. 3).

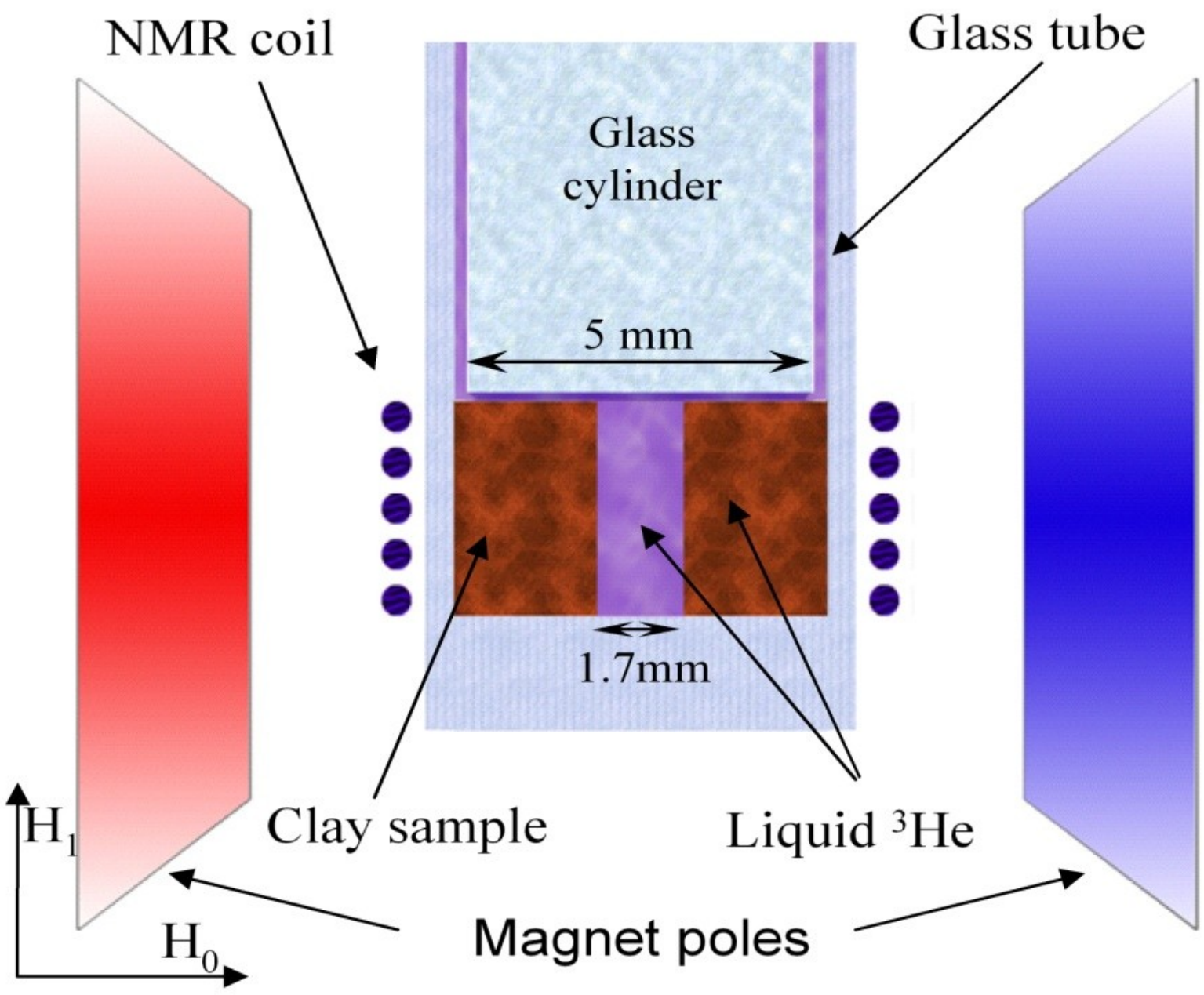

**Figure 3.** Geometry of the NMR experiments.

The longitudinal magnetization relaxation time $T_1$ of $^3$He was measured by the saturation recovery method using spin-echo signal. The observation of the spin-echo signal was possible due to specially induced inhomogeneity of the external magnetic field $H_0$ (NMR probe has been decentered in the magnet).

## 3. Results and discussion

The nuclear spin relaxation time $T_1$ of liquid $^3$He in clay sample was measured on frequency 12 MHz at the temperature 1.5 K. The recovery of the spin echo signal is non single-exponential, what can be seen on Fig. 4. The liquid $^3$He nuclei spin echo signal recovery curve can be described by the following equation with the following fitting parameters: $A_0 = 0.014 \pm 0.004$ arb.units, $A = 1.004 \pm 0.007$ arb.units, $T_1 = 7.609 \pm 0.130$ s, $n = 0.626 \pm 0.008$.:

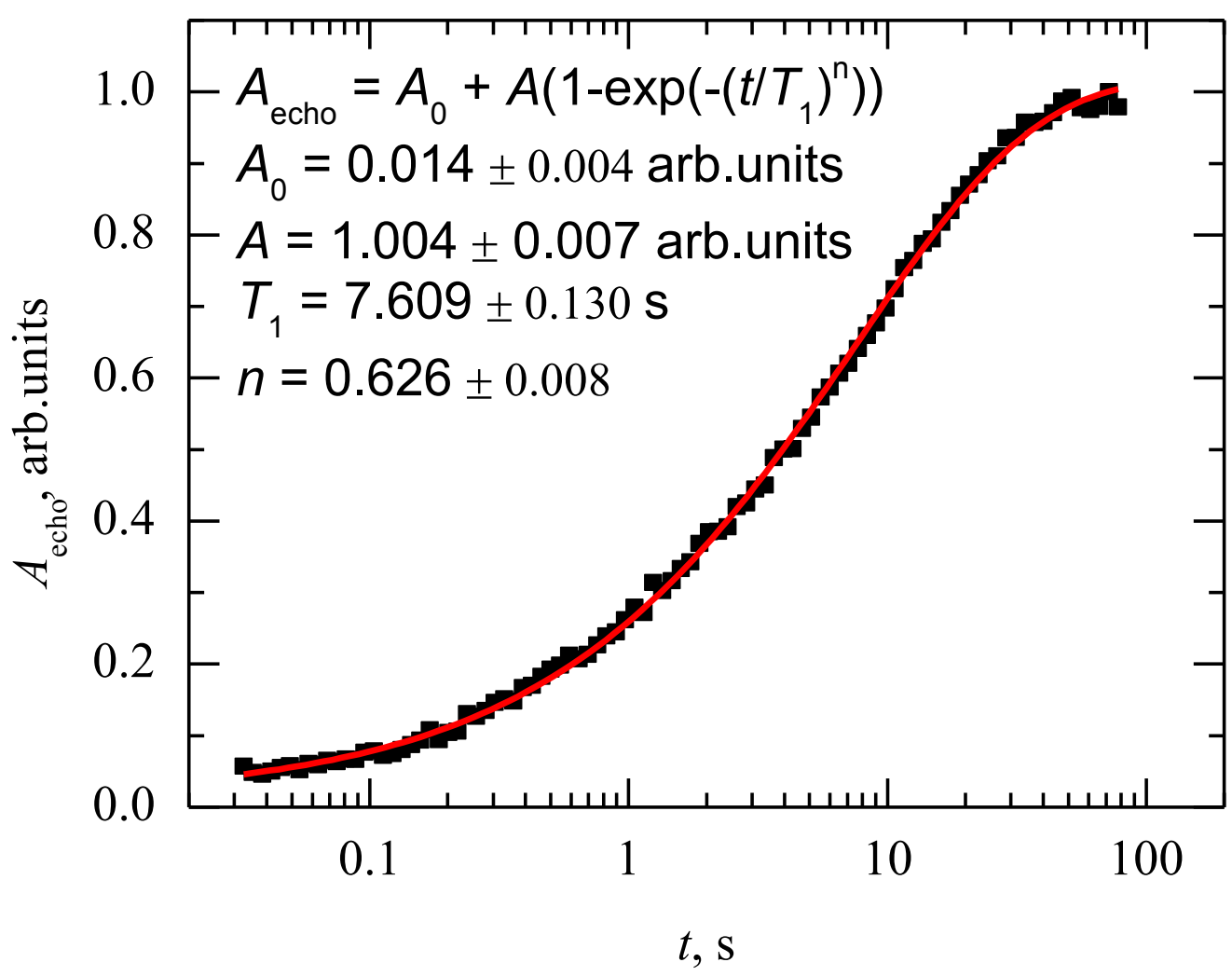

**Figure 4.** The recovery of liquid $^3$He nuclei spin echo signal in clay sample on frequency 12 MHz at temperature 1.5 K (■ – experimental data).

$$A_{echo} = A_0 + A\left(1 - \exp\left(-\left(\frac{t}{T_1}\right)^n\right)\right) \qquad (5)$$

The power $n = 0.626 \pm 0.008$ in the equation (5) indicates the presence of distribution of relaxation times $T_1$. The curve contains information about all components but it is necessary to inverse data to obtain it.

The distribution of relaxation times computed from real experimental data by UPEN are shown on Fig. 5. We see narrow peak (in semi-log scale) and wide peak with shorter $T_1$ values. The narrow peak corresponds to $^3$He in the calibration cavity. Compared this distribution with mono-exponential distributions (Fig. 2) we can assume that the relaxation in the calibration cavity is mono-exponential, as well. The wide peak corresponds to $^3$He relaxation in the pores of clay sample.

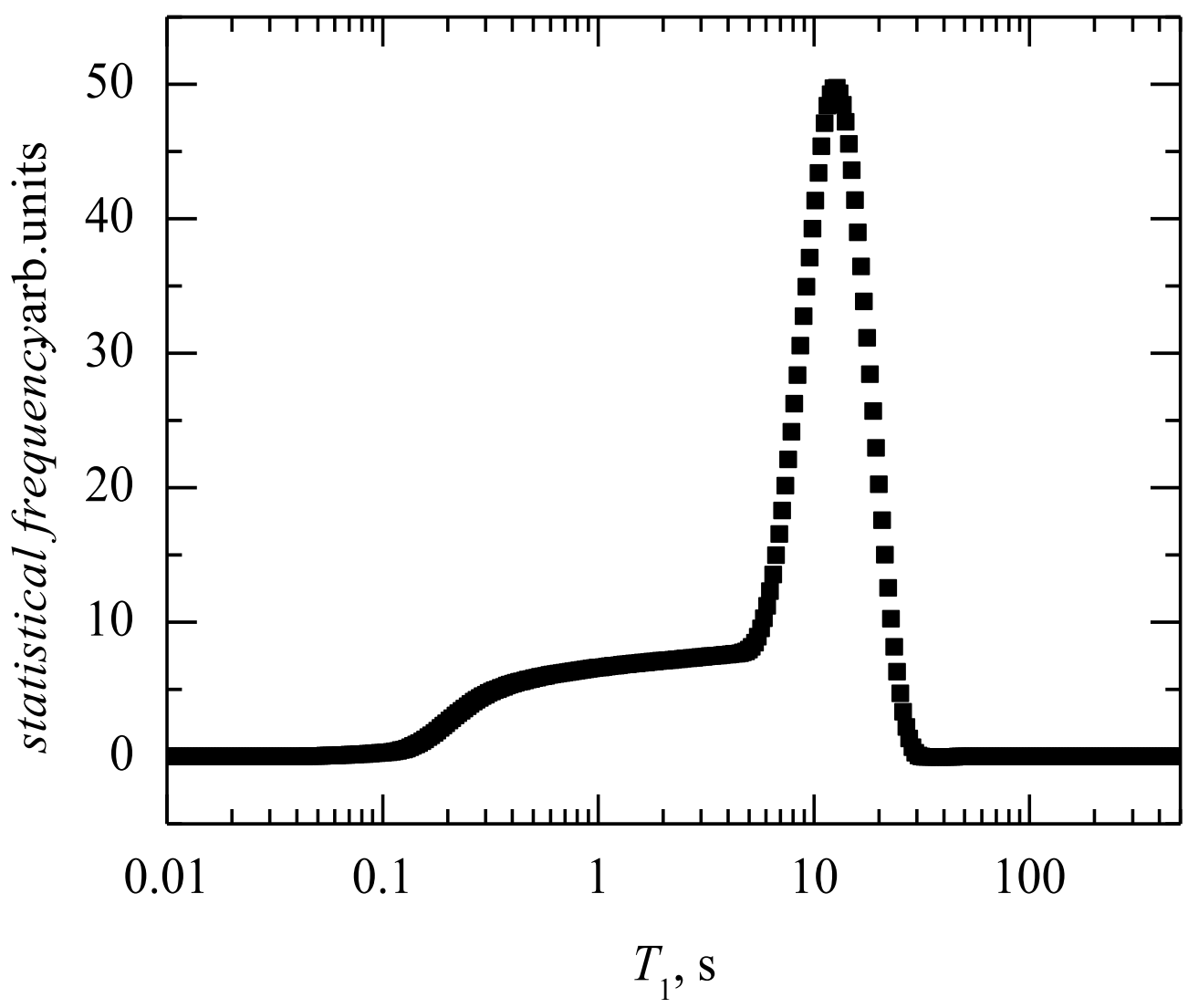

**Figure 5.** Distribution of liquid $^3$He $T_1$ relaxation times for a clay sample obtained by regularized inverse Laplace transform UPEN algorithm.

Thus, inverse Laplace transform revealed unique information about our system, which was not shown on the relaxation curve (Fig. 4). According to analysis of relaxation curve our system should have distribution of $T_1$, which was indicated by power $n \approx 0.6$ in the equation (5). While, inverse Laplace transform unambiguously shows the presence of two different areas of our system (calibration cavity and clay pores), where relaxation differs significantly.

The distribution of relaxation times, obtained by inverse Laplace transform, could be converted into pore size distribution, using specific models of relaxation. For instance, in case of very small pore's sizes, when the molecular diffusion of $^3$He during the time of an experiment is longer than the size of pores the longitudinal relaxation of liquid $^3$He would be single-exponential in each pore. The $T_1$ relaxation mechanism in this case would be surface relaxation, and direct knowledge of surface relaxation time $T_{1s}$, would allow to obtain pore's size distribution. The surface relaxation itself could have different mechanisms, specially complicated by using geological sample, like in our case. Certainly, in some specific cases, the NMR $^3$He relaxation in each pore could be non single-exponential and in these cases inverse Laplace transform would not allow to obtain pore's size distribution.

## 4. Summary

We have demonstrated that the UPEN algorithm of the inverse Laplace transform of $^3$He relaxation curves in porous media can give information about properties of media. The distribution of relaxation times could be converted into pore size distribution, using specific models of relaxation.

## Acknowledgments

We wish to acknowledge A.V. Egorov for enlightening discussions.